\documentstyle[epsfig]{EuroPhys}
\input EuroMacr
\begin{document}
%
%
%
\euro{29}{6}{1-$\infty$}{1997}
\Date{12 June 1997}
\shorttitle{A. P. JAUHO and Ned S. WINGREEN: PHASE MEASUREMENT   ETC.}
%
%
%
\title{Phase measurement of photon-assisted tunneling 
through a quantum dot}
\author{A. P. Jauho\inst{1} \And Ned S. Wingreen\inst{2}}
\institute{
     \inst{1} Mikroelektronik Centret, Technical University of Denmark, Bldg 345east\\
DK-2800 Lyngby, Denmark\\
     \inst{2} NEC Research Institute, 4 Independence Way, Princeton, NJ 08540
}
%
%
\rec{12 June 1997}{in final form 12 June 1997}
%
%
%
\pacs{
\Pacs{73}{20.Dx}{Electron states in low-dimensional structures (superlattices, quantum well structures
and multilayers)}
\Pacs{73}{23.Hk}{Coulomb blockade; single-electron tunneling}
\Pacs{73}{23.$-$b}{Mesoscopic systems}
\Pacs{73}{40.Gk}{Tunneling}
      }
\maketitle
%
%
%
\begin{abstract}
Recent double-slit interference experiments [Schuster {\it et al.},
Nature {\bf 385} (1997) 417 ] have demonstrated the possibility
of probing the phase of the complex transmission coefficient of
a quantum dot via the Aharonov-Bohm effect.  
We propose an extension of these experiments: an ac
voltage imposed on the side gate with the concomitant
photonic sidebands leads to additional structure both in the
amplitude and in the phase of the Aharonov-Bohm signal.  
Observation of these effects would be a definitive proof
of coherent absorption and reemission of photons from the ac source.
\end{abstract}
\begin{figure}

%
\epsfxsize=7.0cm
\epsfbox{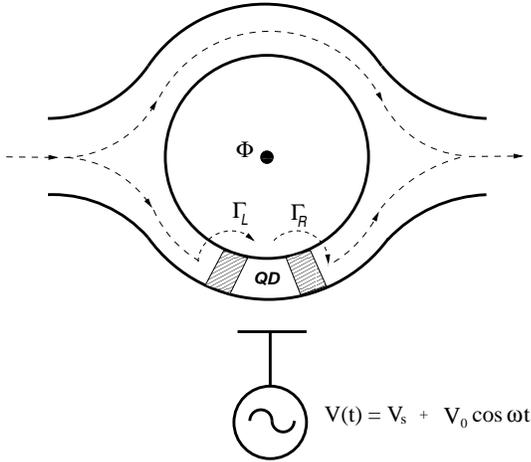}
\caption{Schematic lay-out of the proposed double-slit interference 
experiment. The device consists of an Aharonov-Bohm ring with a
quantum dot (QD) in one arm. A time-dependent voltage $V(t)$ is applied
to the quantum dot via a side gate. 
(Four-terminal measurement is
implied to eliminate the effects of multiple reflections in the ring.)}
\label{f1}
\end{figure}

%
%
%
%
Phase coherence is the hallmark of all mesoscopic transport
phenomena.  Yet normal transport measurements
yield information only about the magnitude of the transmission amplitude,
and not its phase.  In a groundbreaking set of experiments, 
Yacoby {\it et al.} \cite{Yacoby} and Schuster {\it et al.} 
\cite{Schuster} recently
demonstrated that a phase
measurement is nevertheless possible in a mesoscopic double-slit
geometry. Their experimental protocol can be
summarized as follows:
A magneto-transport measurement is performed on an Aharonov-Bohm ring
with a quantum dot fabricated in one of its arms.
If the quantum dot supports coherent transport, the transmission
amplitudes through the two arms interfere. 
A magnetic field induces
 a relative phase change, $2\pi \Phi/\Phi_0$, between the
two transmission amplitudes, $t_0$ and $t_{\scriptscriptstyle{\!Q\!D}}$,
leading to an oscillatory 
conductance $g(B) = (e^2/h){\cal T}(B)$, with 
\begin{equation}
{\cal T}(B)=
{\cal T}^{(0)} +
2{\rm Re}\{ t^*_0 
t_{\scriptscriptstyle{Q\!D}} e^{2\pi i \Phi/\Phi_0}\}
+...,
\label{Tring}
\end{equation}
where $\Phi$ is the flux threading the ring, $\Phi_0=hc/e$ is the flux quantum, 
and where the ellipsis represents higher harmonics due to multiple 
reflections. In the experiments, an oscillatory magnetoconductance of this form
was clearly observed thus demonstrating coherent transmission through
the dot \cite{Yacoby,Schuster}. Furthermore, controlling
the potential on the dot 
with a side-gate voltage
(see fig.\ \ref{f1} for a schematic lay-out),
allowed measurement of the 
phase shift of the transmission amplitude  
$t_{\scriptscriptstyle{Q\!D}}$ through the dot \cite{Schuster}.  
The success of these experiments 
gave rise to a number of other works which 
concentrated on refining  the
interpretation of the experimental results
\cite{Levy,Yacoby2,Hackenbroich,Bruder}. 
Yet, the experiments also suggest  application
to other phase-coherent transport 
processes. One particular example   which has been of considerable recent
interest, both experimentally  
\cite{Guim,Leo,Keay,Zeuner,Ooster} 
and 
theoretically
\cite{Sokol,BS,Wagner,Gloria,Stoof,Stafford},
is photon-assisted tunneling. 
While photon-assisted tunneling (PAT) is intrinsically a 
coherent phenomenon, existing measurements
of PAT are insensitive to the phase of the transmitted electrons
and do not directly demonstrate coherence in the presence of
the time-dependent field. Here we propose a measurement of
photon-assisted tunneling through a quantum dot 
in the mesoscopic double-slit geometry described above. 
In essence, we propose a combination of the experiments of
Kouwenhoven {\it et al.} \cite{Leo,Ooster} 
where a microwave modulated side-gate voltage gave rise
to photon-assisted tunneling through a quantum dot, and the
interference experiments 
of refs. \cite{Yacoby} and \cite{Schuster}.

For an experiment
of this type, we calculate the coherent transmission 
amplitude through the quantum dot in the presence of an arbitrarily
strong ac potential applied to the side gate.
Our theoretical results
indicate that phase-coherent absorption and reemission of photons
can be unambiguously demonstrated via phase measurements at the
sidebands of the main transmission
resonance. In addition, for large driving amplitudes the
phase shift associated with the main transmission resonance
can be reversed from its usual behavior, providing a
direct demonstration of coherence in a strong ac potential.

We focus on transport in the neighborhood of a single Coulomb
oscillation peak associated with a single nondegenerate electronic
level of the quantum dot \cite{Meirav}. The effect of the ac side-gate voltage is
described entirely through the time-dependent energy of this level
\begin{equation}
\epsilon(t)=\epsilon_0(V_{\rm s})+V_{\rm ac}\cos \omega t, 
\label{energy}
\end{equation}
where the static energy of the level $\epsilon_0$ depends on the dc side-gate
voltage $V_{\rm s}$.
 All other levels on the
dot can be neglected provided  the ac amplitude, $V_{\rm ac}$,
 and the photon energy, $\hbar \omega$, are small
compared to the level spacing on the dot.

In the absence of an ac potential, a suitable model for 
the transmission amplitude 
$t_{\scriptscriptstyle{Q\!D}}(\epsilon)$ 
through the dot is the Breit-Wigner form,
\begin{equation}
t_{\scriptscriptstyle{Q\!D}}(\epsilon)={{-i\sqrt{\Gamma_L \Gamma_R}} \over 
{\epsilon - \epsilon_0(V_{\rm s}) + i\Gamma/2}}\;,
\label{tqdstatic}
\end{equation}
where 
$\Gamma=\Gamma_L+\Gamma_R$
is the full width at half maximum of the resonance 
on the dot due to tunneling to the left and right leads.
 Eq. (\ref{tqdstatic}) implies 
a continuous phase accumulation  of $\pi$ 
in the transmission amplitude as the Coulomb
blockade peak is traversed. (Note that 
the Breit-Wigner form is exact for a noninteracting system with $\Gamma$
independent of energy.)

In the dynamic case, the simple Breit-Wigner description
must be generalized, and the object to evaluate
is the $S$-Matrix element \cite{WJW,us}. 
Provided interactions in the leads can be neglected, the 
elastic transmission amplitude $t_{\scriptscriptstyle{Q\!D}}(\epsilon)$
can be written as the energy conserving part
of the $S$-Matrix between the left lead and the right lead
\begin{equation}
\lim_{\epsilon' \rightarrow \epsilon} 
{\langle \epsilon', R | {\cal S} | \epsilon, L \rangle} =
\delta(\epsilon' - \epsilon) t_{\scriptscriptstyle{Q\!D}}(\epsilon).
\label{Svst}
\end{equation}
The $S$-Matrix is simply related to the retarded Green function
of the level on the dot, including both tunneling to the leads and the ac
potential \cite{WJW}
\begin{equation}
{\langle \epsilon', R | {\cal S} | \epsilon, L \rangle} =
-i {{\sqrt{ \Gamma_L \Gamma_R}} \over {2 \pi}}
 \!\!\int\!\!\!\!\int\!\! dt dt_1 e^{i(\epsilon' t - \epsilon t_1)}
G^r(t, t_1).
\label{SvsG}
\end{equation}
Combining eqs.~(\ref{Svst}) and (\ref{SvsG}) allows us to write
\begin{equation}
t_{\scriptscriptstyle{Q\!D}}(\epsilon) = -i \sqrt{\Gamma_L \Gamma_R} 
\langle A(\epsilon,t)\rangle_t   ,
\label{tvsA}
\end{equation}
where the brackets denote a time average, and where
\begin{equation}
A(\epsilon,t) = \int\!\! dt_1 
e^{i\epsilon(t-t_1)} G^r(t,t_1)\;.
\label{A}
\end{equation}
For the time-dependent energy level given by eq. (\ref{energy}), 
we find \cite{us}
\begin{equation}
G^r(t,t_1)=-i\theta(t-t_1)\exp\left[
- {{\Gamma} \over {2}}(t-t_1)
-i\int_{t_1}^t dt' \epsilon(t') \right]
\label{gr}
\end{equation}
so that
\begin{equation}
\langle{ A(\epsilon,t)}\rangle_t = \sum_{k=-\infty}^{\infty}
{{J_k^2(V_{\rm ac}/\hbar \omega)}
\over \epsilon - \epsilon_0(V_{\rm s}) - k\hbar\omega + i\Gamma/2}
\;.
\label{Aave}
\end{equation}

A combination of eqs.~(\ref{tvsA}) and (\ref{Aave}),
evaluated at the Fermi energy, gives
the relevant transmission amplitude, and hence the amplitude of the 
Aharonov-Bohm oscillations at $T=0$ K. 
At finite temperatures one must compute
${t}_{\scriptscriptstyle{Q\!D}}=\int\! 
d\epsilon(-\partial f_0/\partial \epsilon)
{t}_{\scriptscriptstyle{Q\!D}}(\epsilon)$ where $f_0(\epsilon)$ is the Fermi function, 
and the final result is
\begin{equation}
{t}_{\scriptscriptstyle{Q\!D}} =\left(- {\Gamma\over 4\pi T}\right)
\sum_{k=-\infty}^{\infty} J_k^2(V_{\rm ac}/\hbar\omega)
\psi'[{1\over 2}-{i\over 2\pi T}(\mu-\epsilon_0(V_{\rm s})-
k\hbar\omega+i{\Gamma\over 2})]\;,
\label{mainresult}
\end{equation}
where $\psi'$ is the derivative of the digamma function, and
$\mu$ is the chemical potential in the leads.
\begin{figure}
\epsfxsize=8.0cm
\epsfbox{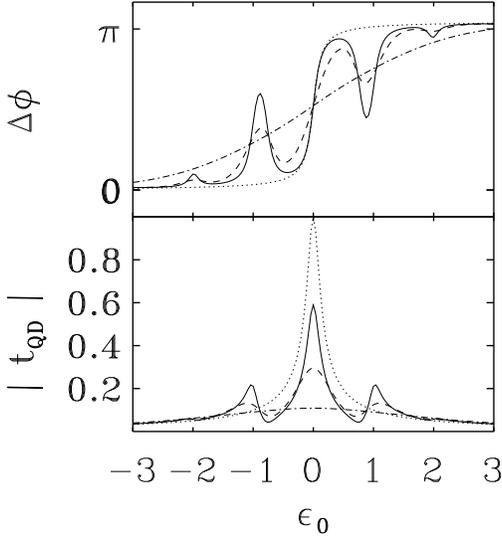}
\vspace{1.5cm}
\caption{Temperature dependence of the phase shift $\Delta\phi$ (top panel) and
the square of the amplitude (bottom) of 
${t}_{\scriptscriptstyle{Q\!D}}$. 
The level-width is $\Gamma/2=0.1$, in terms of which the other
parameters are $V_{\rm ac}=1.0$, $\omega = 1.0$, and 
$T = 0$ (solid line), $0.1$ (dashed line),  
$0.5$ (dash-dotted line).  For comparison, the $T=0$
time-independent results are shown as dots, cf. eq.(3).  } 
\label{f3}
\end{figure}

Eq.(\ref{mainresult}) is the main result of this paper, and in what
follows we shall evaluate it in several cases of interest.  We emphasize
that a conventional conductance measurement would yield information
only about the time average of
{\it the square} of the
transmission amplitude, and the double-slit geometry  
is necessary in order to probe the phase.
Figure \ref{f3} shows the computed magnitude of 
${t}_{\scriptscriptstyle{Q\!D}}$ (bottom)
and its phase (top), as a function of the level energy
$\epsilon_0(V_{\rm s})$.
As compared to the time-independent case (shown as a dotted line),
several features are noteworthy.  The magnitude of
${t}_{\scriptscriptstyle{Q\!D}}$ 
shows photonic side-bands, reminiscent of those seen in transmission
through a microwave modulated quantum dot \cite{Leo}.  
However, there is an important difference from the usual case of photon-assisted
tunneling. The amplitude of the Aharonov-Bohm oscillation is 
sensitive only to the time average of 
the transmission amplitude
$t_{\scriptscriptstyle{Q\!D}}$.
Hence only elastic transmission through the dot contributes,
{\it i. e.}, the net number of photons absorbed from the ac
field must be zero. The sideband at say $\epsilon = \epsilon_0(V_s) -
\hbar \omega$ corresponds to a process in which an electron
first absorbs a photon to become resonant at energy $\epsilon_0(V_s)$,
and subsequently reemits the photon to return to its original energy.
The requirement that the net photon absorption be zero also satisfies
the quantum mechanical dictate that interference is only possible
if no trace is left of the passage of the electron through the 
quantum dot.

\begin{figure}
\epsfxsize=8.0cm
\epsfbox{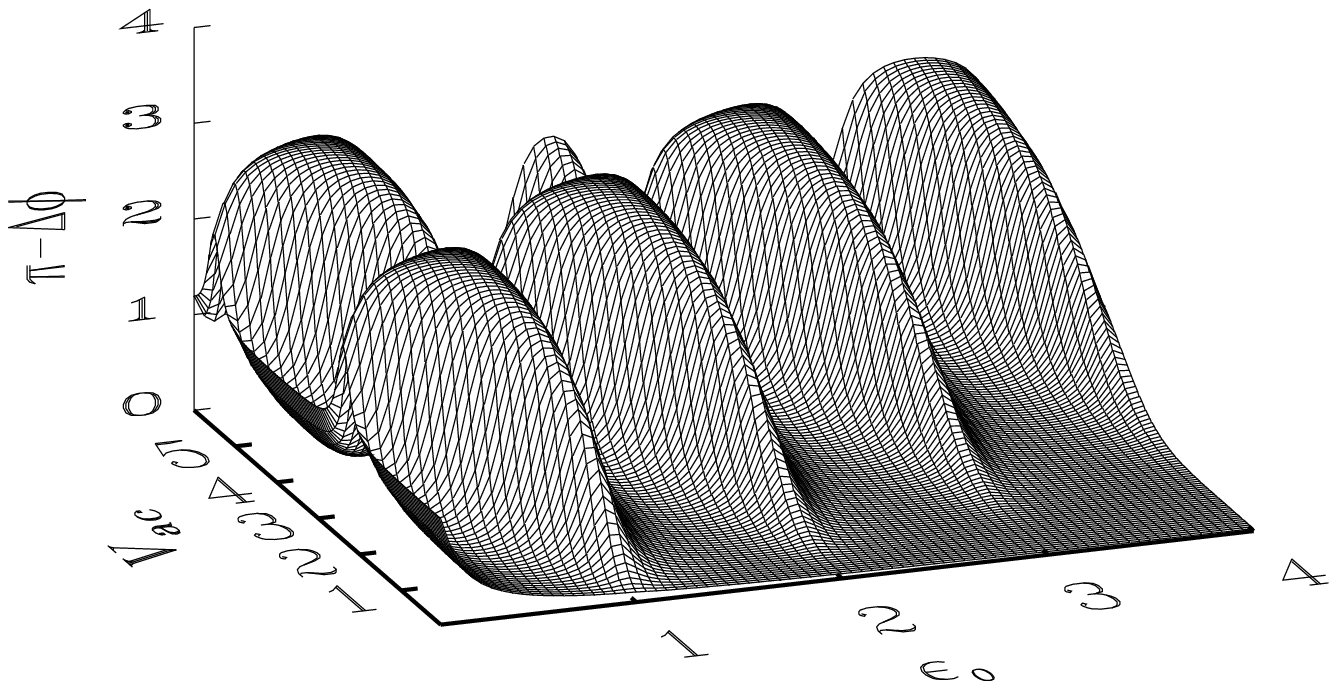}
\vspace{.5cm}
\epsfxsize=8.0cm
\epsfbox{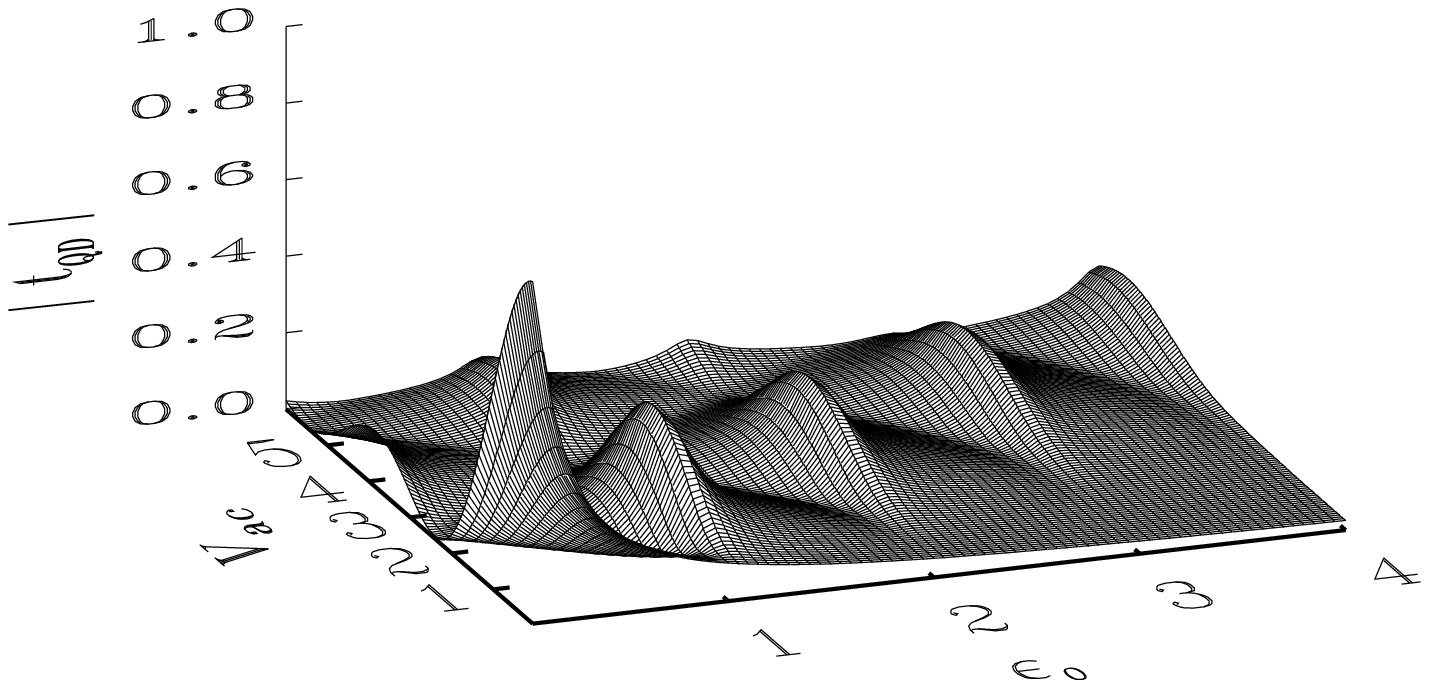}
\caption{The phase shift $\pi-\Delta\phi$
(top panel) and the square of the amplitude (bottom)
for $\omega=1.0, \Gamma/2 = 0.1, T=0$.
The energy axis corresponds to $\epsilon_0(V_s)$ with
$\mu=0$.} 
\label{f2}
\end{figure}

Perhaps most interesting
are the features appearing in the phase: the phase shift shows a
non-monotonic behavior, with pronounced resonances located at the
energies corresponding to the photonic side-bands.  The strengths
of these phase resonances are strongly dependent on the ac amplitude
$V_{\rm ac}$, and in fig.\ \ref{f2} we show
the computed signal as a function of both $\epsilon_0(V_{\rm s})$ and the 
amplitude of modulation.
\begin{figure}
\epsfxsize=8.0cm
\epsfbox{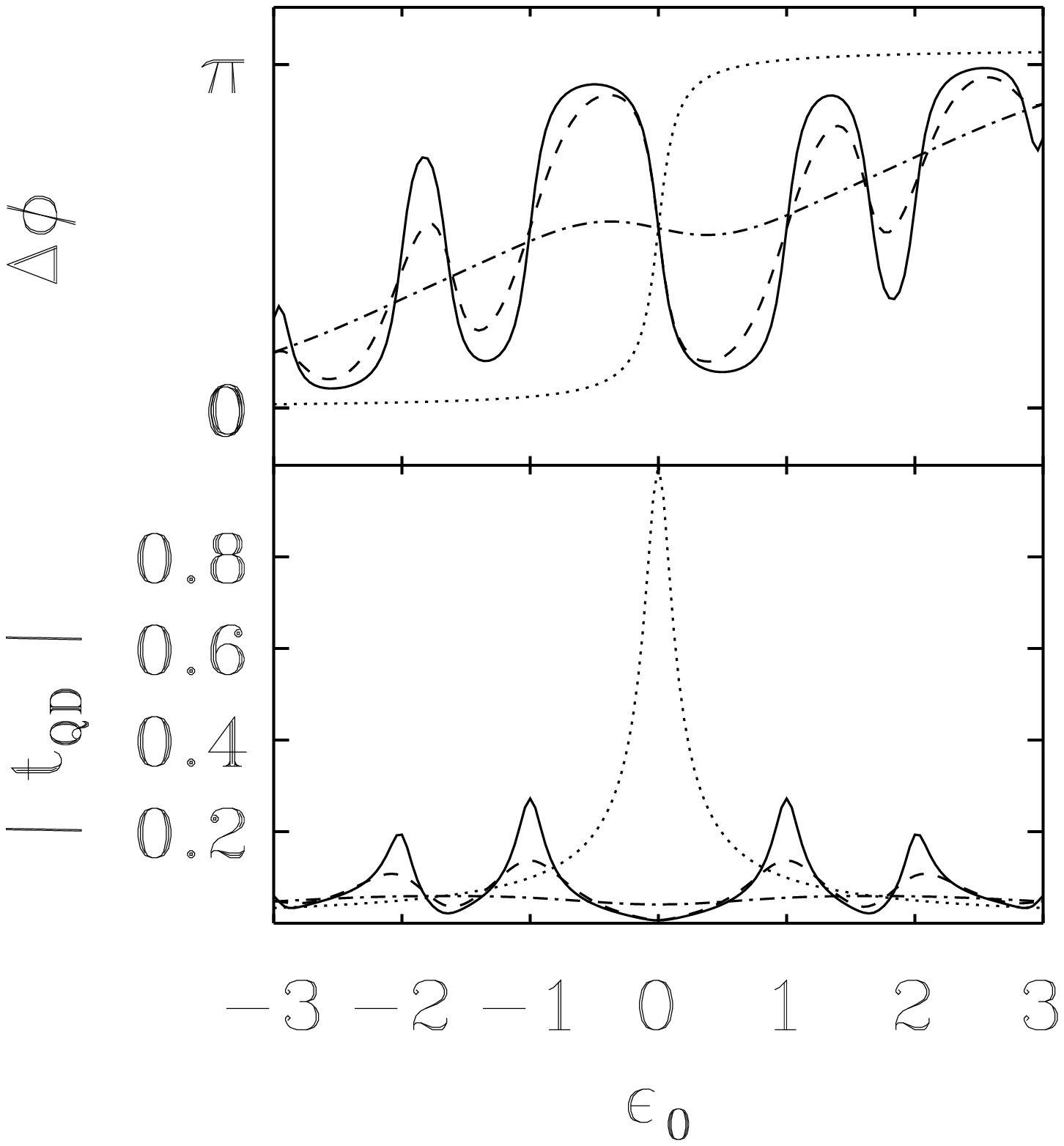}
\vspace{1.5cm}
\caption{Temperature dependence of the phase shift (top panel) and
the amplitude (bottom) of 
${t}_{\scriptscriptstyle{Q\!D}}$, for
$V_{\rm ac}=2.405$, $\omega=1.0$, $\Gamma/2=0.1$,
$T = 0, 0.1, 0.5$, with linetypes as in fig. 2.  Note the qualitative differences as
compared to fig. 2: suppression of the main transmission
peak, and the negative
slope of the phase shift at $\epsilon_0=0$.}
\label{f4}
\end{figure} 
In fig.\ \ref{f4} we highlight another interesting consequence of
eq.(\ref{mainresult}): it is possible to entirely {\it quench} the
main transmission peak (bottom panel), or {\it change the sign} of
the slope of the phase at resonance by adjusting the ratio $V_{\rm ac}/\hbar\omega$
to coincide with a zero of the Bessel function $J_0$ (top).  This phenomenon is mathematically
analogous to the recently observed absolute negative conductivity
in THz-irradiated superlattices \cite{Keay}; in our case, however,
it is the {\it phase} rather than the current that displays this behavior.

In summary, we have proposed an experiment to probe phase-coherence
in a quantum dot driven by a strong ac potential. The phase
measurement relies on the mesoscopic double-slit geometry pioneered
in refs. \cite{Yacoby} and \cite{Schuster}. The amplitude of
Aharonov-Bohm oscillations reflects the amplitude for coherent
transmission through the dot with zero net absorption of photons.
We find that coherent absorption and reemission of photons can
be unambiguously detected via phase measurement at sidebands of
the main transmission resonance through the quantum dot.

\stars
The authors acknowledge useful comments from
Karsten Flensberg, Ben Yu-Kuang Hu, and Andreas Wacker. 

%
%


\vskip-12pt
\begin{thebibliography}{99}
%
\bibitem{Yacoby}
\Name{A. Yacoby, M. Heiblum, D. Mahalu, \And H. Shtrikman}
\Review{Phys. Rev. Lett.} 
\Vol{74} \Year{1995} \Page{4047}.

\bibitem{Schuster}
\Name{R. Schuster, E. Buks, M. Heiblum, D. Mahalu, V. Umansky,
\And H. Shtrikman} 
\Review{Nature} \Vol{385} \Year{1997} \Page{417}.

\bibitem{Levy}
\Name{A. Levy Yeyati and M. B{\"u}ttiker} 
\Review{Phys. Rev. B} \Vol{52}  \Year{1995} \Page{R14360}.

\bibitem{Yacoby2}
\Name{A. Yacoby, R. Schuster, \And M. Heiblum}
\Review{Phys. Rev. B} \Vol{53} \Year{1995} \Page{9583}.
 
\bibitem{Hackenbroich}
\Name{G. Hackenbroich And H. A. Weidenm{\"u}ller} 
\Review{Phys. Rev. Lett.} 
\Vol{76} \Year{1996}
\Page{110}; 
\Review{Phys. Rev. B} 
\Vol{53} \Year{1996} \Page{16379}.

\bibitem{Bruder}
\Name{C. Bruder, R. Fazio, \And H. Schoeller} 
\Review{Phys. Rev. Lett.} \Vol{76}  \Year{1996} \Page{114}.


\bibitem{Guim}
\Name{P. S. S. Guimar{\~a}es, B. J. Keay, J. P. Kaminsky, S. J. Allen, Jr.,
P. F. Hopkins, A. C. Gossard, L. T. Florez, \And J. P. Harbison}
\Review{Phys. Rev. Lett.} \Vol{70} \Year{1993} \Page{3792}.

\bibitem{Leo}
\Name{L. Kouwenhoven, S. Jauhar, J. Orenstein, P. L. McEuen,
Y. Nagamune, J. Motohisa, \And H. Sakaki}
\Review{Phys. Rev. Lett.} \Vol{73} \Year{1993} \Page{3443}.

\bibitem{Keay}
\Name{B. J. Keay, S. Zeuner, S. J. Allen, Jr., K. D. Maranowski, A. C. Gossard,
U. Bhattacharya, \And M. J. W. Rodwell}
\Review{Phys. Rev. Lett.} \Vol{75} \Year{1995} \Page{4102}.

\bibitem{Zeuner}
\Name{S. Zeuner, B. J. Keay, S. J. Allen, Jr., K. D. Maranowski, A. C. Gossard,
U. Bhattacharya, \And M. J. W. Rodwell}
\Review{Phys. Rev. B} \Vol{53} \Year{1996} \Page{1717}.

\bibitem{Ooster}
\Name{T. H. Oosterkamp, L. P. Kouwenhoven, A. E. A. Koolen, N. C. van der Vaart,
\And C. J. P. M. Harmans}
\Review{Phys. Rev. Lett.} 
\Vol{78} \Year{1997} \Page{1536}.

\bibitem{Sokol}
\Name{D. Sokolovski} 
\Review{Phys. Rev. B} 
\Vol{37} \Year{1988} \Page{4201}.

\bibitem{BS}
\Name{C. Bruder \And H. Sch{\"o}ller}
\Review{Phys. Rev. Lett.} \Vol{72} \Year{1994} \Page{1076}.

\bibitem{Wagner}
\Name{M. Wagner}
\Review{Phys. Rev. Lett} \Vol{76} \Year{1996} \Page{4010}.

\bibitem{Gloria}
\Name{J. I{\~n}arrea \And G. Platero}
\Review{Europhys. Lett.} \Vol{34}  \Year{1996} \Page{43}.

\bibitem{Stoof}
\Name{T. H. Stoof \And Yu. V. Nazarov} 
\Review{Phys. Rev. B} \Vol{53}  \Year{1996} \Page{1050}.

\bibitem{Stafford}
\Name{C. A. Stafford \And N. S. Wingreen}
\Review{Phys. Rev. Lett.} \Vol{76} \Year{1996} \Page{1916}.

\bibitem{Meirav}
\Name{U. Meirav \And E.B. Foxman}
\Review{Semicond. Sci. Technol.} \Vol{10} \Year{1995} \Page{255}.


\bibitem{WJW}
\Name{N. S. Wingreen, K. W. Jacobsen, \And J. W. Wilkins}
\Review{Phys. Rev. B} \Vol{40} \Year{1989} \Page{11834}.

\bibitem{us}
\Name{N. S. Wingreen, A. P. Jauho, \And Y. Meir}
\Review{Phys. Rev. B} \Vol{48} \Year{1993} \Page{8487};
\Name{A. P. Jauho, N. S. Wingreen, \And Y. Meir}
\Review{Phys. Rev. B} \Vol{50} \Year{1994} \Page{5528}.
\end{thebibliography}
\end{document}